\documentclass[preprint,12pt]{elsarticle}

\usepackage{amssymb}
\usepackage{subcaption}
\usepackage{times}
\usepackage{epsfig}
\usepackage{graphicx}
\usepackage{amsmath}
\usepackage{amssymb}
\usepackage{float}
\usepackage{makecell}
\usepackage{color}
\usepackage{multirow}
\usepackage{graphicx,placeins}
\usepackage[T1]{fontenc}
\newcommand{\latinphrase}[1]{\textit{#1}} 
\newcommand{\etal}{\latinphrase{et~al.}\xspace}

\usepackage{csquotes}
\usepackage{graphicx}
\usepackage{soul}
\usepackage{float}
\usepackage{multirow}
\DeclareUnicodeCharacter{0301}{\'{e}}
\usepackage{amsmath}
\usepackage{times}
\usepackage{epsfig}
\usepackage{graphicx}
\usepackage{amsmath}
\usepackage{amssymb}
\usepackage{float}
\usepackage{makecell}
\usepackage{color}
\usepackage{multirow}
\usepackage{graphicx,subcaption,placeins}
\usepackage[T1]{fontenc}
\usepackage{xcolor}         
\usepackage{multirow}
\usepackage{color, colortbl}
\usepackage{subcaption}
\usepackage{bm}

\usepackage{hyperref}
\usepackage{graphics}
\usepackage{graphicx}
\usepackage{url}

\renewcommand{\arraystretch}{2}
\definecolor{Gray}{gray}{0.9}
\definecolor{LightCyan}{rgb}{0.88,1,1}
\colorlet{lightgray}{gray!30}
\colorlet{heavygray}{gray!60}
\usepackage[normalem]{ulem}
\usepackage{xspace}
\def\arraystretch{1.2}%  1 is the default, change whatever you need

\usepackage{times}
\usepackage{epsfig}
\usepackage{graphicx,subcaption}
\usepackage{amsmath}
\usepackage{amssymb}
\usepackage{lipsum}
\usepackage{mathtools, nccmath}

\usepackage{amsmath}

\usepackage[linesnumbered,ruled,vlined]{algorithm2e}
\PassOptionsToPackage{hyphens}{url}\usepackage{hyperref}

\usepackage[cmintegrals]{newtxmath}
\makeatother
\begin{document}

\begin{frontmatter}

\title{KAN-LSTM: Benchmarking Kolmogorov-Arnold Networks for Cyber Security Threat Detection in IoT Networks}

\author[inst1]{Mohammed Hassanin}
% \author[inst1,inst2]{Mohammed Hassanin}
% \author[inst3]{Ibrahim Radwan}
% \author[inst1]{Maryam Ghodrat}

\affiliation[inst1]{organization={University of New South Wales Canberra (UNSW Canberra)},%Department and Organization
            state={ACT},
            country={Australia}}

% \affiliation[inst2]{organization={Fayoum University},%Department and Organization 
%             state={Fayoum},
%             country={Egypt}}
% \affiliation[inst3]{organization={University of Canberra},%Department and Organization 
%             state={ACT},
%             country={Australia}}

\begin{abstract}

By utilising their adaptive activation functions, Kolmogorov-Arnold Networks (KANs) can be applied in a novel way for the diverse machine learning tasks, including cyber threat detection. KANs substitute conventional linear weights with spline-parametrized univariate functions, which allows them to learn activation patterns dynamically, inspired by the Kolmogorov-Arnold representation theorem. In a network traffic data, we show that KANs perform better than traditional Multi-Layer Perceptrons (MLPs), yielding more accurate results with a significantly less number of learnable parameters. We also propose KAN-LSTM model to combine advantages of spatial and temporal encoding. The suggested methodology highlights the potential of KANs as an effective tool in detecting cyber threats and offers up new directions for adaptive defensive models. Lastly, we conducted experiments on three main dataset, UNSW-NB15, NSL-KDD, and CICID2017, as well as we developed a new dataset combined from IOT-BOT, NSL-KDD, and CICID2017 to present a stable, unbiased, large-scale dataset with diverse traffic patterns. The results show the superiority of KAN-LSTM and then KAN models over the traditional deep learning models.  The source code is available at \textcolor{blue}{\href{https://github.com/mfawzy/KAN-LSTM-Benchmarking-Kolmogorov-Arnold-Networks-for-Cyber-Security-Threat-Detection-in-IoT}{GitHub repository}}.

\end{abstract}

\end{frontmatter}

\section{Introduction}
Machine learning has greatly benefited from the development of neural networks in all the paradigm of science, ranging from tabular data such as cyber security \cite{hassanin2024comprehensive},  to multidimensional ones including images, videos and 3D \cite{hassanin2025crossformer3d}. The Multi-Layer Perceptron (MLP)  is a fundamental architecture in this field that is comprised of numerous layers of completely connected nodes \cite{popescu2009multilayer}. In several disciplines, this structure has shown to be useful for approximating nonlinear functions \cite{hornik1989multilayer}. 

MLPs often struggle with the challenges posed by large-scale and relational datasets because of their structure, which is not well-suited to capturing spatial patterns in the data. This causes issues with scalability and efficiency, and as a result, researchers have focused on developing architectures that can better handle such complexities. Lately, Convolutional Neural Networks (CNNs) \cite{lecun2002gradient} have become a dominant neural network model for processing various types of data. This is because CNNs introduce modifications that allow them to more effectively capture the intricate structures present in input data. These networks are designed to automatically learn spatial and relational hierarchies of features through backpropagation [15], [16], making them highly effective for most of the tasks such as textual, visual, and 3D tasks.

The recent introduction of Kolmogorov-Arnold Networks (KANs) \cite{kiamari2024gkan, bodner2024convolutional} offers a novel approach to traditional neural network architectures by utilizing learnable functions on the graph edges, rather than fixed activation functions at the nodes. This design is based on the Kolmogorov-Arnold Representation Theorem \cite{schmidt2021kolmogorov}, which asserts that any multivariate continuous function can be broken down into univariate functions and a summation. In KANs, each layer applies a distinct, learnable function to its inputs. Liu et al.  argued that KANs can overcome some inherent limitations of MLPs, particularly in managing complex functional mappings in high-dimensional data \cite{liukan}.

The use of KANs neural structures for cyber threat detection opens up a new area of interest. KANs are providing notable benefits due to the structured nature of cyber security data and the critical requirement for precise function approximation in threat detection and classification. ConvKANs \cite{bodner2024convolutional}, a convolutional adaption of KANs, have demonstrated encouraging results in early trials by substituting a learnable nonlinear activation for the conventional dot product in pattern recognition. By utilizing Kolmogorov-Arnold theorem to identify sophisticated attacks in high-dimensional security data,  better machine learning patterns will be able to detect emergent cyber threats.

Several experiments are in progress to examine how KANs might improve machine learning behaviors in different fields \cite{cheon2024demonstrating}.  KAN-GPT project \cite{GANESH2024KANGPT} applies KANs to Generative Pre-trained Transformers (GPTs) for language-based analysis. Furthermore, utilising KANs in GPT-2 models instead of MLPs.

While there is increasing interest in creating distinct KAN variants for use in many domains \cite{bresson2024kagnns, jamali2024learn, nagai2024kolmogorov}, a formal assessment of KANs' applicability for threat detection tasks is still missing. In order to bridge this gap, this study offers preliminary empirical study of KANs' performance in cyber defence, especially in threat detection. Using cyber security datasets such network traffic and attack simulations, we experiment with concepts from ConvKANs \cite{bodner2024convolutional} implementations to evaluate their performance in terms of accuracy, training efficiency, and model parameters. It is anticipated that our research will clarify the advantages and disadvantages of KANs for cyber defence suitability, particularly detection of cyber attacks.

\begin{figure}[]
    \centering
     \includegraphics[width=\linewidth]{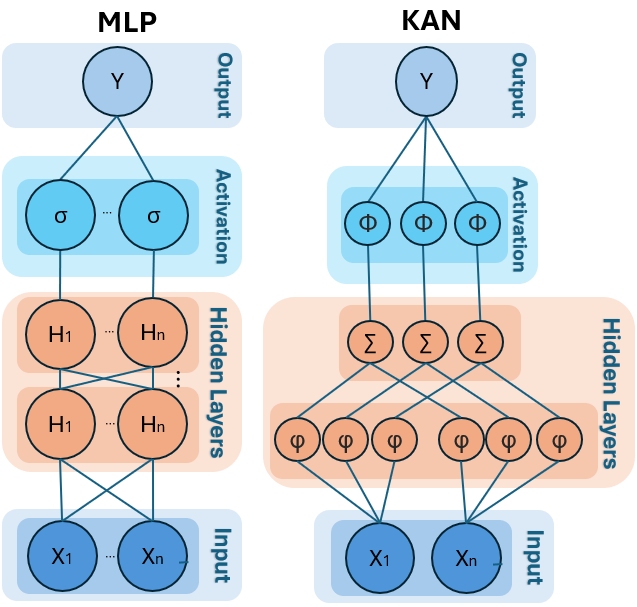}
   \caption{Visual explanation for MLPs and KANs.  }
  \label{fig:KANs}
\end{figure}
\section{THEORETICAL FORMULATION}
This section explores the theoretical and mathematical foundations of KANs \cite{schmidt2021kolmogorov} and how they are modified for use in CNNs. Learning parametric functions on graph edges, a technique applicable to both convolutional layers and conventional fully-connected layers in neural networks, is the fundamental breakthrough of KANs. Within this context, convolutional networks that integrate KAN ideas are referred to as "ConvKAN" in general. To specifically highlight the effectiveness of KAN-based convolutional layers in increasing cyber threat identification and response, we designate to the network as "KConvKAN" when both convolutional and linear layers are implemented with KANs. Figure 1 provides visual description to KANs and the modified version used in this paper.

\subsection{Multi-Layer Perceptrons (MLPs)}

A fully connected feedforward neural network made up of several layers of nodes (neurons) is called a Multi-Layer Perceptron (MLP). By connecting to every other node in the subsequent layer, each node in a layer applies a nonlinear activation function to the weighted sum of its inputs. The Universal Approximation Theorem \cite{cybenko1989approximation}, which states that, given appropriate activation functions, a feedforward network with a single hidden layer and a finite number of neuron may approximate any continuous function over compact subsets of ${R^n}$, is the main source of support for the popularity of MLPs. An MLP with L layers can be expressed mathematically as follows:

\begin{equation}
\small  \text{MLP}(x) = \sigma_L \left( W_L \sigma_{L-1} \left( W_{L-1} \cdots \sigma_1 \left( W_1 x + b_1 \right) \cdots  + b_{L-1} \right) + b_L \right),
\end{equation}

where $\sigma$ denotes the activation function at layer $i$, $\mathbf{W}_i$ represents the weight matrix for layer $i$, and $\mathbf{b}_i$ is the bias vector for layer $i$.

\subsection{Kolmogorov-Arnold Networks (KANs)}
Here, we present the ideas of Liu et al. [1] to introduce the KANs principles. For further information, interested readers are directed to [1].

\textbf{Kolmogorov-Arnold Representation Theorem:} 

The Kolmogorov-Arnold Representation Theorem [32] asserts that any multivariate continuous function can be expressed as a finite composition of continuous functions of a single variable, combined with the binary operation of addition, especially for a smooth function $f: \mathbb{R}^n \to \mathbb{R} $

\[
f(x) = \sum_{i=1}^{2k+1} \phi_i \left( \sum_{j=1}^k \psi_{ij}(x_j) \right)
\]
where 
$phi_i$
  and 
$\psi_{ij}$
  are continuous functions of a single variable, and 
$k$ is a finite integer. This representation allows complex multivariate functions to be decomposed into simpler univariate components, which facilitates their approximation and modeling through networks like KANs.

\textbf{KAN Architecture}
KolmogorovArnold representation is generalized by KANs via using learnable activation functions on graph edges. A KAN layer with $k_{in}$-dimensional inputs and $k_{out}$-dimensional outputs is defined as

\[
\Phi = \left\{ \phi_{i,i} \mid j = 1,2,\ldots,k_{\text{in}}, \; i = 1,2,\ldots,k_{\text{out}} \right\}
\]

where $\phi_{i,i}$ are learnable functions, parameterized as splines. The output of a KAN layer is as follows:

\[
x_{l+1}^j = \sum_{i=1}^{k_l} \phi_{l,j,i}(x_{l,i}), \quad j = 1, \ldots, k_{l+1}
\]

or it can be expressed in matrix form for easy understanding as follows:

\[
x_{l+1} = \Phi_l x_l
\]
where $\Phi_l$ is the function matrix corresponding to the $l^{th}$ KAN layer.
More precisely, the approximation bound of KANs follows:
\[
\left\|f - \left(\Phi_{G}^{L-1} \circ \cdots \circ \Phi_{G}^{0}\right) x\right\|_{C^m} \quad C_m \leq CG^{-r-1+m}
\]

where $G$ is the grid size, $r$ is the order of the B-spline, and $C$ is a constant as defined by~\cite{ref1}, where $\Phi_{G}^{L-1} \circ \cdots \circ \Phi_{G}^{0}$ defines the composition of the function matrices from layers $0$ to $L-1$, whereas the composition operation $\circ$ indicates that the output of one function matrix is used as the input of the next.

\section{Related Works}
The advancement in machine learning due to deep learning architectures has influenced all the paradigms of computer science, including cyber security and machine learning \cite{hassanin2024visual, hassanin2025pllm}. This progressed from MLPs \cite{almahadeen2024enhancing}, CNNs \cite{moustafa2022dfsat}, and LSTM \cite{yuan2019attention} networks to Transformers \cite{de2024vincent} have been extensively studied for intrusion and threat detection in publicly published datasets such as NSL-KDD, CICIDS2017, and BOT-IoT.

MLPs are among the earliest neural architectures applied to intrusion detection. Shone \etal \cite{shone2018deep}  introduced a stacked MLP-based autoencoders for NSL-KDD. This achieved competitive accuracy over traditional machine learning methods. Mirza and Cosan \cite{mirza2018computer} explored an MLP architecture for CICIDS2017 that  improved detection of low-frequency attacks. However, MLPs often struggle to capture spatial and temporal dependencies inherent in network traffic data.

CNNs have been used to capture spatial correlations among network features. Yin \etal applied 1D-CNNs to CICIDS2017 traffic which gained high precision in DDoS detection \cite{yin2017deep}. Li \etal transformed network traffic features into 2D representations to improve detection of probe and U2R attacks \cite{li2015hybrid}. Kim \etal developed a CNN-based intrusion detection system for IoT networks \cite{kim2017method}.

Since LSTMs used to model temporal dependencies, it has been employed to detect abnormalities in sequential traffic data. Early, Staudemeyer and Omlin \cite{staudemeyer2015applying} showed that LSTMs can extract long-range dependencies in network traffic data compared to  traditional RNNs. Hybrid approaches like CNN-LSTM have been proposed to combine CNN’s spatial feature extraction and LSTM’s temporal modeling \cite{halbouni2022cnn, deore2022hybrid, abdallah2021hybrid}.

Other works highlight the importance of deep learning ensembles. \cite{ye2025daily} used attention-based models to encode threat detections representations. \cite{ferrag2020deep} presented a comparative study of CNN, LSTM, and MLP models for intrusion detection across NSL-KDD, CICIDS2017, and UNSW-NB15 datasets.  This concluded that hybrid architectures generally outperform standalone models.

Overall, threat detection in IoT networks have been witnessing progression due to the rapid advancements of deep learning models. However, KANs are different in the essence of providing non-linear activation functions which can be a milestone for a new era of deep learning progression. For cyber security tasks, this study provides a comprehensive comparative study on the three popular dataset in the field as a benchmark for the next studies. Also, this study proposes a hybrid model, called KAN-LSTM, to combine the advantages of non-local (KANs) and local (LSTM) representation encoding. In other words, to encode temporal relationships and long-ranges ones.

\begin{figure*}[]
\renewcommand{\arraystretch}{1.5}
    \centering
     \includegraphics[width=\linewidth]{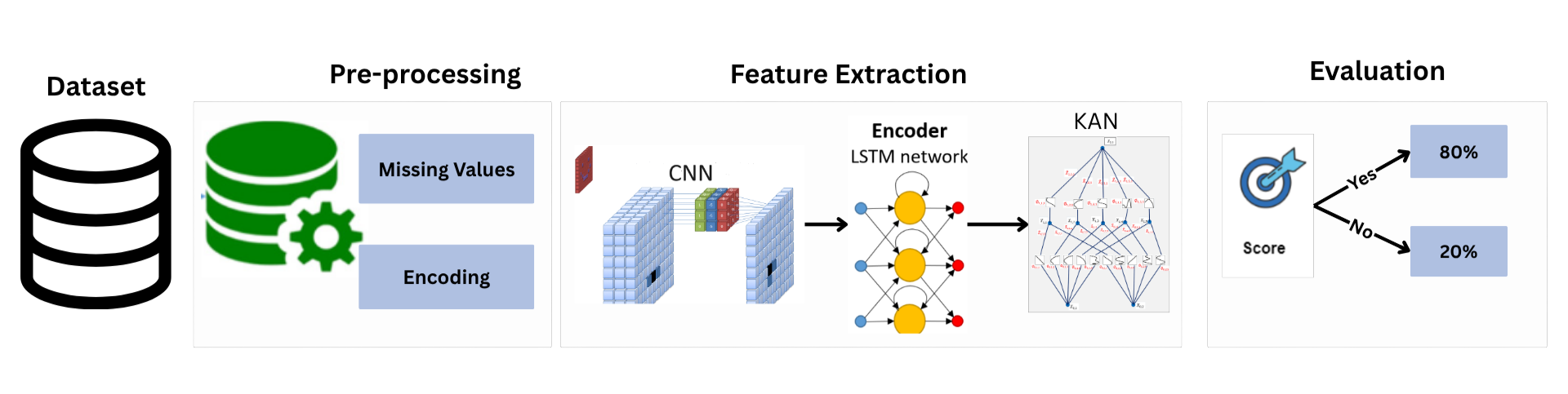}
   \caption{Visual explanation for KAN-LSTM architecture. It is composed of two layers of CNN, followed by one layer of LSTM, and finally one KAN layer. This after the stages of pre-processing.}
  \label{fig:proposed}
\end{figure*}

\section{Method}
As Figure \ref{fig:proposed} shows,  the proposed method consists of three main stages: Convolution Neural Network (CNN), Long Short-Term Memory (LSTM), KANs. This architecture is mainly proposed to combine the benefits of LSTM in handling sequence data and KAN to encode non-local representation in a learning approach. Also, it mitigate the problem of LSTM to retain the learned keypoints in a large-scale and the issue of KANs to encode local descriptive.

\subsubsection{Convolutional Neural Network (CNN)}
The CNN captures local feature dependencies. For an input tensor $x$, a convolutional layer computes:
\begin{equation}
h_{i,j}^{(l)} = \sigma\left( \sum_{m=1}^{M} w_m^{(l)} \cdot x_{i+m, j} + b^{(l)} \right)
\end{equation}
where $w_m^{(l)}$ is the kernel, $b^{(l)}$ the bias, and $\sigma$ an activation function (e.g., ReLU).

In KAN-LSTM, two CNN layers have been employed to encode the local relationships between the traffic input data. While it is known that CNN is able to encode features of 2D tensors better than 1D ones, we transformed the 1D input into 2D tensors. 
\subsubsection{LSTM Encoder}
CNN output is passed to a LSTM network to capture temporal and sequential dependencies. For input sequence $(\mathbf{z}_1, \dots, \mathbf{z}_T)$:
\begin{align}
f_t &= \sigma(W_f [h_{t-1}, \mathbf{z}_t] + b_f) \\
i_t &= \sigma(W_i [h_{t-1}, \mathbf{z}_t] + b_i) \\
o_t &= \sigma(W_o [h_{t-1}, \mathbf{z}_t] + b_o) \\
\tilde{c}_t &= \tanh(W_c [h_{t-1}, \mathbf{z}_t] + b_c) \\
c_t &= f_t \odot c_{t-1} + i_t \odot \tilde{c}_t \\
h_t &= o_t \odot \tanh(c_t)
\end{align}
The final encoder state $h_T$ serves as a compact representation. KAN-LSTM is employing only one LSTM layer to just encode the temporal dependencies amongst the network traffic input. 

\subsubsection{KAN}
The LSTM output is fed into a KAN to model nonlinear high-dimensional mappings. For $m$ inputs $\mathbf{u} = [u_1, \dots, u_m]$ and $n$ outputs:
\begin{equation}
y_j = \sum_{i=1}^m a_{ij} \phi_{ij}(u_i), \quad j=1,\dots,n
\end{equation}
where $a_{ij}$ are trainable weights and $\phi_{ij}(\cdot)$ are learnable spline functions.

\subsection{Loss Function}
Finally, the output of KAN is passed to a classifier. For binary classification,  Binary Cross-Entropy (BCE) loss is used to measure the difference between the predicted probability and the true binary label. It is defined as:
\begin{equation}
\mathrm{BCE} = -\frac{1}{N} \sum_{i=1}^N \left[ y_i \log(p_i) + (1 - y_i) \log(1 - p_i) \right]
\end{equation}
where $N$ is the number of observations, $y_i \in \{0,1\}$ is the actual binary label of the $i$-th observation, $p_i \in [0,1]$ is the predicted probability of the $i$-th observation belonging to class $1$.

\section{Experiments}

\section{Results}
This part covers experimetal evaluations of KANs and KAN-LSTM on the network datasets, such as UNSW-NB-15 \cite{moustafa2015unsw}, NSL-KDD, CICID2017, Tri-IDS. These datasets are chosen because their large scale, diversity, domain coverage for most of the cyber security paradigms . Experiment settings, baselines, datasets, and assessment measures are specified in the following parts. Subsequently, to emphasise the significance of KANs including the proposed model in relation to the baselines, quantitaive and qualitative comparisons with the most advanced techniques are presented.

\subsection{Experimental Setup}
In the backend, implementation models are Pytorch, cuDNN, and CUDA-11. In the training phase, same parameters are set to ensure equitable comparisons .  AdamW \cite{loshchilov2017decoupled} uses the same configurations for all the methods. With $32$GB, the Quadro GV100 server is in use. The learning rate is $2e-5$, whereas the number of epochs is $30$.  No transfer learning or fine-tuning of any of models.

\subsection{Datasets}
This study is benchmarking the use of KANs performance on cyber security datasets.The purpose of this selection was to assess KAN's performance on real-world datasets, which are extensively utilised in the industry. We choose the datasets taking into account many factors including testing KAN in the scenario of plenty and  few features. Here are the details of the considered datasets:
\begin{itemize}
    \item \textbf{UNSW\_NB15}
     dataset was developed at the Australian Centre for Cyber Security \cite{moustafa2015unsw}. It contains a hybrid of real and simulated network traffic and is widely used for intrusion detection research. The dataset covers various attack categories, including DoS, worms, backdoors, and exploits, and provides 49 features extracted from network flow data.

 \item \textbf{NSL-KDD}
   is an enhanced version of the KDD Cup 1999 dataset, which was originally used for network intrusion detection research \cite{tavallaee2009detailed}. NSL-KDD removes redundancies and imbalance issues present in the original dataset, making it more effective for training and testing machine learning models. It includes four types of attacks: DoS, U2R, R2L, and Probe.

\item \textbf{CICIDS2017}
   created by the Canadian Institute for Cybersecurity, CICIDS2017 is a comprehensive dataset for network intrusion detection \cite{sharafaldin2018toward}. It features realistic traffic patterns generated by simulating normal and malicious user behavior, covering a wide range of attacks like DDoS, brute force, and infiltration. The dataset consists of labeled data and 80 features.

 \item \textbf{Tri-IDS Results}
   is a dataset that that is combined of three popular datasets, BOT-IOT \cite{koroniotis2019towards}, NSL-KDD, CICID, coined as Tri-IDS. IOT-BOT simulates normal and malicious Internet of Things (IoT) network traffic. It covers a variety of IoT attacks such as denial of service (DoS), distributed denial of service (DDoS), and information theft, helping to advance research in IoT security. However, BOT-IOT includes a lot of noisy records as we wanted to provide a large-scale diverse dataset to be a strong benchmark to evaluate the various datasets, especially these datasets have been witnessing 100\% accuracy.
   \end{itemize}

These datasets present a variety of difficulties, including a range of classification problems, missing values, and varying amounts of occurrences and features. Also, we chose them to cover different tasks of cyber threat detections. This versatility makes it possible to thoroughly assess KANs' effectiveness in a variety of real-world circumstances. We provide a summary of the datasets' primary attributes in Table \ref{tab:datasets}.

\begin{table*}[]
    \centering
    \resizebox{\textwidth}{!}{
    \begin{tabular}{lccccc}
    \hline
        Datasetname& Task& NumberofFeatures &NumberRows &Missingvalues& DataType  \tabularnewline
        \hline
        UNSW\_NB15 & Intrusion detection &49&2,5m&None& numerical, categorical \tabularnewline
         NSL-KDD&  Intrusion detection&41&148,517&None& numerical, categorical \tabularnewline
         CICIDS2017& Intrusion detection &80&3M&Few&numerical, categorical  \tabularnewline
         Tri-IDS& IoT security, botnet detection, intrusion detection&120&~76.5 million&few&numerical, categorical, time-series\tabularnewline
         \hline
    \end{tabular}
    }
    \caption{Simple comparison between the involved datasets in the experiments.}
    \label{tab:datasets}
\end{table*}

\subsection{Baselines}
The comparisons of the proposed model along with deep learning models are presented in this section. Simply, CNN, MLP, LSTM, KANs are employed to signify the importance of the proposed model. Furthermore, compared to deep learning models, they are more susceptible to imbalance of the datasets. In recent decades, these methods of learning has earned popularity due to being trained over a predetermined number of epochs to learn problems' patterns.

\subsection{Evaluation}
The study benchmarks KANs by comparing their performance and effectiveness with traditional MLPs, CNNs, LSTMs in threat detection tasks. The aim is to assess whether KANs can be a viable alternative  in cyber security. The publicly published datasets are employed to explore the efficacy of KANs in detecting cyber threats. The benchmarking process evaluates both models across these datasets with maintaining the fairness amongst the baselines. This balanced setup illustrates which model offers superior performance under equivalent conditions. 
\subsection{Model Configuration}
To evaluate performance, we used a variety of KAN, ConvKAN,  MLP, LSTM, KAN-LSTM.
\textbf{Baseline Models:} 
 \begin{itemize}
     
 \item \textbf{CNNs}
 To create a point of comparison for assessing the KANs and ConvKANs, we employed conventional MLPs and CNNs as baselines. 
\item \textbf{CNNs:}  
We implemented CNNs as a strong baseline for spatial feature extraction in learning models. The CNNs consisted of three convolutional layers with ReLU activations, followed by pooling layers and two MLP layers. This to capture local patterns that makes it a standard reference point for evaluating the performance of KAN-based models,  as well as, CNNs have been extensively used in almost all the previous threat detection studies. 

\item \textbf{LSTM:}  
LSTM networks were used as a sequence modeling baseline to address the vanishing gradient problem via gated memory units and capture temporal relationships. LSTM showed significance in time-series in general and cyber threat detection in special. One layer is used followed by MLP layer, not bidirectional setting.

\item \textbf{KANs:}  
KANs replace fixed activation functions in traditional MLPs with learnable univariate functions. As KANs suffer from stability, scalability and slow processing, we chose this repository to amend such difficulties \cite{Blealtan2024}.  

\item \textbf{ConvKAN:}  
ConvKAN integrates the KANs with a convolutional frameworks. Three ConvKAN layer along with three KAN layers are used together to build the model.

\item \textbf{KAN-LSTM:}  
Figure \ref{fig:proposed} details the settings of the proposed model. Simply put, two CNN layers, one LSTM layer, and two KAN layers. The details KAN-LSTM is a hybrid architecture where KAN-based layers perform adaptive feature extraction before feeding the processed features into an LSTM for temporal modeling. 

\end{itemize}
\subsection{Evaluation protocols} 
To evaluate the proposed approach, KAN-LSTM, we employ various criteria, such as recall, precision, F-score $(F_1)$, and precision. The following is a detailed explanation of them:

\textbf{Accuracy:} 
is among the fundamental performance metrics. In contrast to incorrect predictions $T_n$, the accurate predictions are primarily denoted as $T_p$. The proportion of accurate samples to the total is called accuracy.

\begin{equation}
    Acc = \frac{T_p + T_n}{T_p + T_n + F_p + F_n}
\end{equation}

\textbf{Precision}
The score of the true predicted samples $T_p$ in relation to the accurate ones is computed using this measure. 
\begin{equation}
    Prec = \frac{T_p}{T_p +  F_p }
\end{equation}

\textbf{Recall:}
In performance evaluation, the recall measure is typically used to determine the number of real positive predictions made by the model. Determining the attack detection rates is crucial.

\begin{equation}
    Rec = \frac{T_p}{T_p +  F_n }
\end{equation}

\textbf{F-measure ($F_1$)} is a metric used to weigh the advantages of recall and precision. As a trade-off between recall and precision, $F_1$ is more significant than accuracy, especially when there are many negatives in a class with an uneven distribution.

\begin{equation}
    F-measure = 2 \times \frac{Prec \times Rec}{Prec +  Rec }
\end{equation}

% \textbf{False Negative Rate (FNR)} The ratio of false negatives to total positives, which includes both false negatives and genuine positives.

% \begin{equation}
%     FNR = \frac{F_n}{F_n + T_p}
% \end{equation}

%  \textbf{Receiver Operating Characteristic (ROC)} is a visual criterion that compares the true positive rate (TPR) against the false positive rate (FPR) at various threshold values. The following metrics are used on the x- and y-axes, respectively, in the ROC graphic. 
 
% % \begin{equation}
% \begin{align}
%     TPR = \frac{T_p}{ T_p + F_n} \\
%     FPR = \frac{F_p}{F_p + T_n } 
% \end{align}
% % \end{equation}

% \textbf{Area Under the ROC Curve (AUC)}. Being both scale- and classification-threshold-invariant, this measure assesses how effectively the predictions are ranked and how unimportant they are in relation to the threshold of the classification. This makes it an attractive metric.

% \textbf{Matthews Correlation Coefficient (MCC)} The most effective classification metric that takes into account the four metrics—TP, TF, FP, and FN—and summarises the confusion matrix is the {Matthews Correlation Coefficient (MCC)}.

% \begin{equation}
%     MCC = \frac{T_p \times T_n - F_p \times F_n}{\sqrt{(T_p+F_p)(T_p+F_n)(T_n + F_p)(T_n + F_n)} }
% \end{equation}

% Under the configurations above, true positives are denoted by $T_p$, true negatives by $T_n$, false positives by $F_p$, and false negatives by $F_n$.

\begin{table*}[]

\centering
\resizebox{\textwidth}{!}{%
\begin{tabular}{llllllllll}
\cline{1-6} \cline{7-10}
\multirow{2}{*}{Model} & \multicolumn{4}{c}{UNSW-NB15}                           &  & \multicolumn{4}{c}{NSL-KDD}                             \\ \cline{2-5} \cline{7-10} 
                       & Accuracy                & Precision & Recall & F1 Score &  & Accuracy                & Precision & Recall & F1 Score \\ 
                       
\cline{1-10}
CNN                    & 93.10 &64.70& 66.10&65.43 &  & 91.66& 66.66& 75.00& 75.00           \\
LSTM                   & 93.97 &68.83& 70.28& 69.55 &        & 89.58& 68.75&68.75& 68.75          \\
MLP                    & 92.57& 62.67& 62.88& 62.82           &  & 90.62 &68.75 &33.33 &70.96     \\
% Transfomer             & 97.00 &96.00& 98.00& 97.00         &  & 98.50& 97.80 &98.60 &98.20           \\
% TransKANs              &  &   &    &   &  &     &   &   &          \\
ConvKAN               &94.31& 70.95          &71.91        & 70.41         &  &93.70                       &    80.25       &\textbf{\textcolor{blue}{ 82.15}}       &  \textbf{\textcolor{blue}{81.90  }}      \\
KANs(2)               &  93.91 &  69.55  & 69.58    & 69.56   &  & 91.66     & 75.00    & 75.00   &   75.00  \\
KANs(5)               &  \textbf{\textcolor{blue}{94.45 }}&  \textbf{\textcolor{blue}{72.42 }} & \textbf{\textcolor{blue}{72.76} }  & \textbf{\textcolor{blue}{72.56}}   &  &\textbf{\textcolor{blue}{93.75}}    &\textbf{\textcolor{blue}{81.25 }}    & 81.25   &   81.25  \\
\hline
KAN-LSTM (Ours)   &   \textbf{\textcolor{red}{95.89} }                     &\textbf{\textcolor{red}{94.50 }}          & \textbf{\textcolor{red}{93.20}}       &  \textbf{\textcolor{red}{ 94.01}}      & & \textbf{\textcolor{red}{ 94.05} }                   &\textbf{\textcolor{red}{98.95}}           & \textbf{\textcolor{red}{92.15}  }     &  \textbf{\textcolor{red}{94.35} }       \\  

\hline
\multirow{2}{*}{Model} & \multicolumn{4}{c}{CICIDS2017}                           &  & \multicolumn{4}{c}{Tri-IDS}                             \\ \cline{2-5} \cline{7-10} 
                       & Accuracy                & Precision & Recall & F1 Score &  & Accuracy                & Precision & Recall & F1 Score \\ 
                       
\cline{1-10}
CNN               &82.75&82.75&40.50&43.89&           &92.60&91.50&91.20&  91.90        \\
LSTM               &86.05&86.323&85.68&85.47                 & &93.05&95.21&91.22&93.17     \\

MLP (2)            &97.61 &77.03& 78.10&77.55&              &93.05&\textbf{\textcolor{blue}{94.70}}&90.74&92.68     \\

MLP (5)            &98.43 &77.03&77.07&77.04&&93.10&93.54&92.12&92.83     \\

ConvKAN            &\textbf{\textcolor{blue}{98.55}} &\textbf{\textcolor{blue}{91.48}}&\textbf{\textcolor{red}{99.69}}&\textbf{\textcolor{blue}{95.41}}&&\textbf{\textcolor{blue}{93.45}}&94.12&\textbf{\textcolor{red}{92.50}}&\textbf{\textcolor{blue}{93.36}}     \\

KANs (2)           &97.69 &77.03&76.78&76.89&&\textbf{\textcolor{blue}{93.45}}&94.34&92.01&93.16     \\

KANs (5)           &98.20 &77.03&77.07&77.04&&93.20&92.82&92.76&92.79     \\
\hline
KAN-LSTM (Ours)    &\textbf{\textcolor{red}{99.28} }&\textbf{\textcolor{red}{99.06}} &\textbf{\textcolor{blue}{99.30}} &\textbf{\textcolor{red}{99.66}} &&\textbf{\textcolor{red}{93.76}} &\textbf{\textcolor{red}{95.85}} &\textbf{\textcolor{blue}{91.06}} &\textbf{\textcolor{red}{93.39} }   \\  \hline
\end{tabular}%
}
\caption{Visual performance comparison of various deep learning models for threat detection across four benchmark datasets: UNSW-NB15, NSL-KDD, CICIDS2017, and Tri-IDS (a combined dataset of BOT-IOT + CICIDS2017 + NSL-KDD). The models evaluated include CNN, LSTM, MLP with 2 and 5 hidden layers, Convolutional ConvKAN, KANs with 2 and 5 layers, and the proposed KAN-LSTM hybrid model. For each dataset, the classification metrics reported are Accuracy, Precision, Recall, and F1 Score. Results illustrate that the proposed KAN-LSTM consistently achieves superior performance across all datasets. With high gains in Precision and F1 Score on CICIDS2017 and NSL-KDD, the proposed method indicates strong generalization and balanced detection of both benign and malicious traffic. Tri-IDS results further confirm the robustness of KAN-LSTM against diverse and large-scale IoT-related attack patterns. For easy reading, we highlighted the top two results in each metric in red, blue color for top one and second top. respectively.}
\label{tab:mytable}
\end{table*}
\section{Results}

\subsection{UNSW-NB15 Results}
Table~\ref{tab:mytable} (top left) shows the performance comparison on the UNSW-NB15 dataset. Among the baseline methods, LSTM achieved the highest accuracy (93.97\%), followed by CNN (93.10\%) and MLP (92.57\%). While these models achieved reasonable accuracy, their precision and recall scores illustrate some limitations in handling diverse attack patterns.

For KAN-based architectures,  they consistently outperformed the baselines. KANs(5) attained the best overall performance among the KAN variants, with an accuracy of 94.45\%, precision of 72.42\%, recall of 72.76\%, and an F1 score of 72.56\%. ConvKAN also delivered strong results (94.31\% accuracy, 70.95\% precision, and 71.91\% recall), showing balanced performance across all metrics. Even smaller configurations  Overall, KAN-based models improved accuracy by 1.3–1.9\% over baselines, with higher precision and recall, indicating superior capability in detecting rare and varied attack types. 

However, the proposed method (KAN-LSTM) obtained the highest performance compared to KAN-based architectures and traditional learning methods. It recorded \textbf{95.89\%} with ~1.5\% above the closest variant. It is noticeable that KAN-LSTM got top results in all the metrics with large margin in precision, recall and F1-score which indicates that the proposed method is capable to overcome the inherited imbalance in the datasets. Same thing applies to KANs.

\subsection{CICIDS2017 Results}
The CICIDS2017 results, shown in Table~\ref{tab:mytable} (bottom left), shows that the best-performing baseline was LSTM (86.05\% accuracy, 85.47\% F1 score), followed by CNN (82.75\% accuracy). However, CNN struggled in recall (40.50\%), indicating difficulty in detecting a wide range of attacks. MLP architectures significantly improved accuracy (97.61\% for MLP-2 and 98.43\% for MLP-5) but provided only slightly better values in precision and recall.

KAN-based models achieved both high accuracy and balanced performance. KANs(2) reached 97.69\% accuracy with an F1 score of 76.89\%, while KANs(5) slightly improved accuracy to 98.20\% with a similar F1 score. ConvKAN variant achieved the strongest performance among all models, with 98.55\% accuracy, 91.48\% precision, and an exceptionally high recall of 99.69\%, yielding an F1 score of 95.41\%. These results confirm that KANs effectively handle imbalanced classes and can generalize well to diverse attack patterns in CICIDS2017.

However, the proposed method (KAN-LSTM) obtained the highest performance compared to all the other methods. It recorded \textbf{99.28\%} with ~5\% above the closest variant. This shows the ability of the proposed method to handle local relationships along with non-local ones. Also, it reflects the power of the proposed method to overlook the imbalance of the datasets and diverse patterns in the input.

\subsection{NSL-KDD Results}
As reported in Table~\ref{tab:mytable} (top right), the CNN baseline achieved 91.66\% accuracy (precision: 66.66\%, F1 score: 75.00\%), indicating fair detection ability. LSTM showed slightly lower accuracy (89.58\%) but balanced precision and recall (both 68.75\%). The MLP model achieved 90.62\% accuracy but suffered from a recall drop to 33.33\%, leading to an F1 score of 70.96\%.

The proposed KAN models obtained superior performance. KANs(5) achieved the highest accuracy (93.75\%) and F1 score (81.25\%), while KANs(2) maintained balanced scores (75.00\% precision, recall, and F1) with competitive accuracy (91.66\%). These results highlight the ability of KANs to achieve high detection performance while improving recall—critical for identifying minority and rare attack classes.

Overall, the proposed method (KAN-LSTM) obtained the best performance compared to comparable variants. It recorded \textbf{94.05\%} with ~1.5\% above the closest variant.

\subsection{Tri-IDS Results}
The combined Tri-IDS dataset results, shown in Table~\ref{tab:mytable} (bottom right), indicate strong baseline performance. CNN achieved the highest accuracy (94.60\%) among baselines, with precision and recall exceeding 94\%. LSTM followed closely with 93.05\% accuracy and balanced precision (95.21\%) and recall (91.22\%). MLP variants delivered similar results, with accuracy around 93\% and minor trade-offs between recall and F1 score.

Our KAN-LSTM model achieved the best overall performance, with 93.76\% accuracy, 95.85\% precision, 91.06\% recall, and a 93.39\% F1 score—surpassing all baseline methods. This consistent improvement across datasets confirms the suitability of KAN-based architectures for complex intrusion detection scenarios, especially those involving heterogeneous traffic patterns and highly imbalanced attack distributions.

KAN-LSTM obtained the highest performance compared to KAN-based architectures and traditional learning methods. It recorded \textbf{93.76\%} with a slight margin of  ~.3\% above the closest variant. This reflects the diversity of this combined dataset as well as the ability of the proposed method to generalize to the pattern diversity. Looking at the results of KAN-LSTM and other methods and how the difference in all the metrics is so slight show the significance of such a combined dataset. It strikes the imbalance in the other datasets. 

Overall, KAN-LSTM is the best candidate in all the comparisons and the datasets. This proves our claim that KAN-LSTM is capable to encode local and non-local relationships.

\begin{figure*}[]
    \centering
    % Row 1: Grouped bar chart (left) & Radar chart (right)
    \begin{subfigure}[t]{0.49\textwidth}
        \centering
        \includegraphics[width=\linewidth]{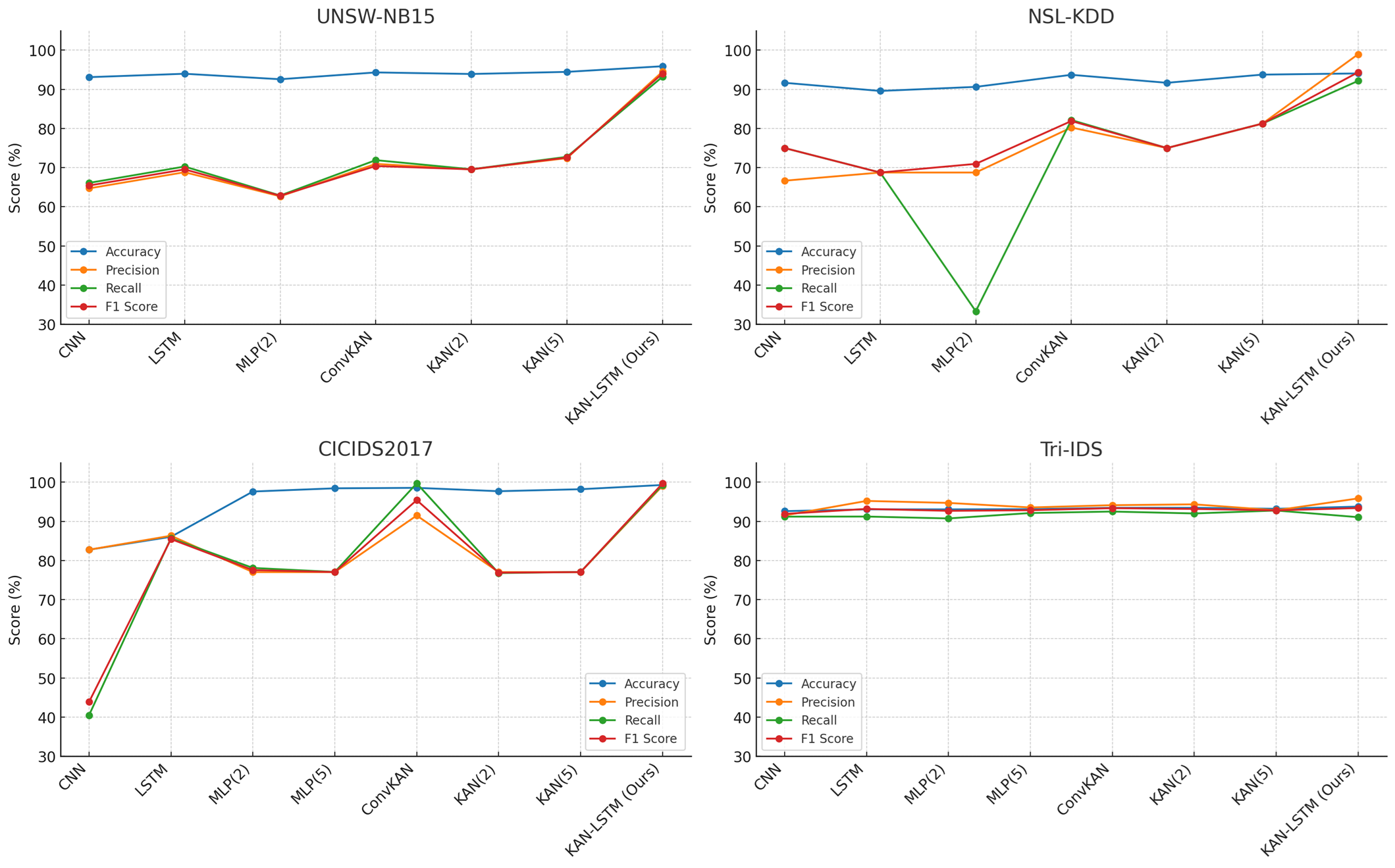}
        \caption{Line Charts.}
        \label{fig:performance_grouped}
    \end{subfigure}
    \hfill
    \begin{subfigure}[t]{0.49\textwidth}
        \centering
        \includegraphics[width=\linewidth]{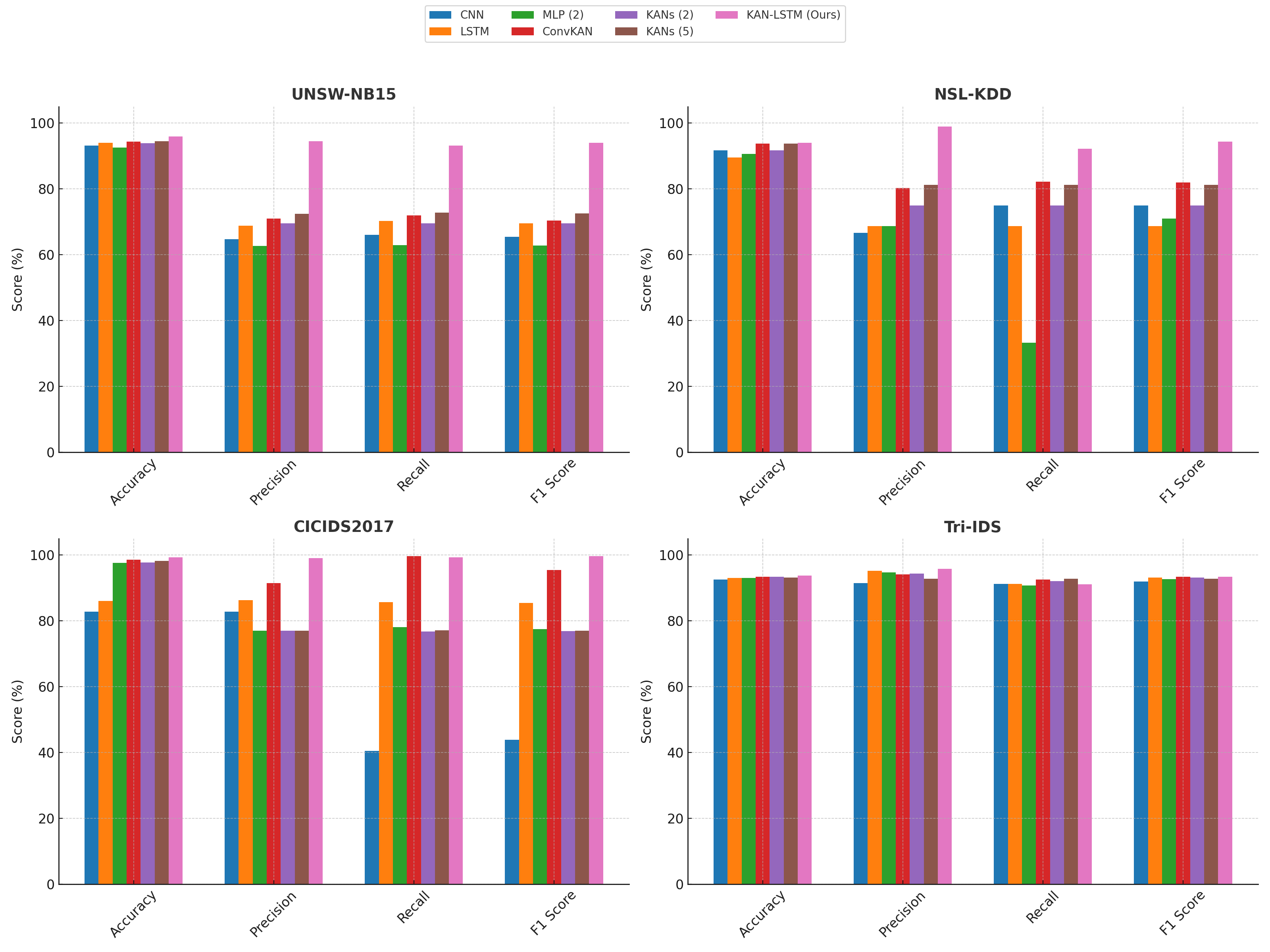}
        \caption{Bar charts.}
        \label{fig:performance_radar}
    \end{subfigure}
    
    % Row 2: Heatmaps (left) & (Optional) You can add another chart or combined table
    \begin{subfigure}[t]{0.49\textwidth}
        \centering
        \includegraphics[width=\linewidth]{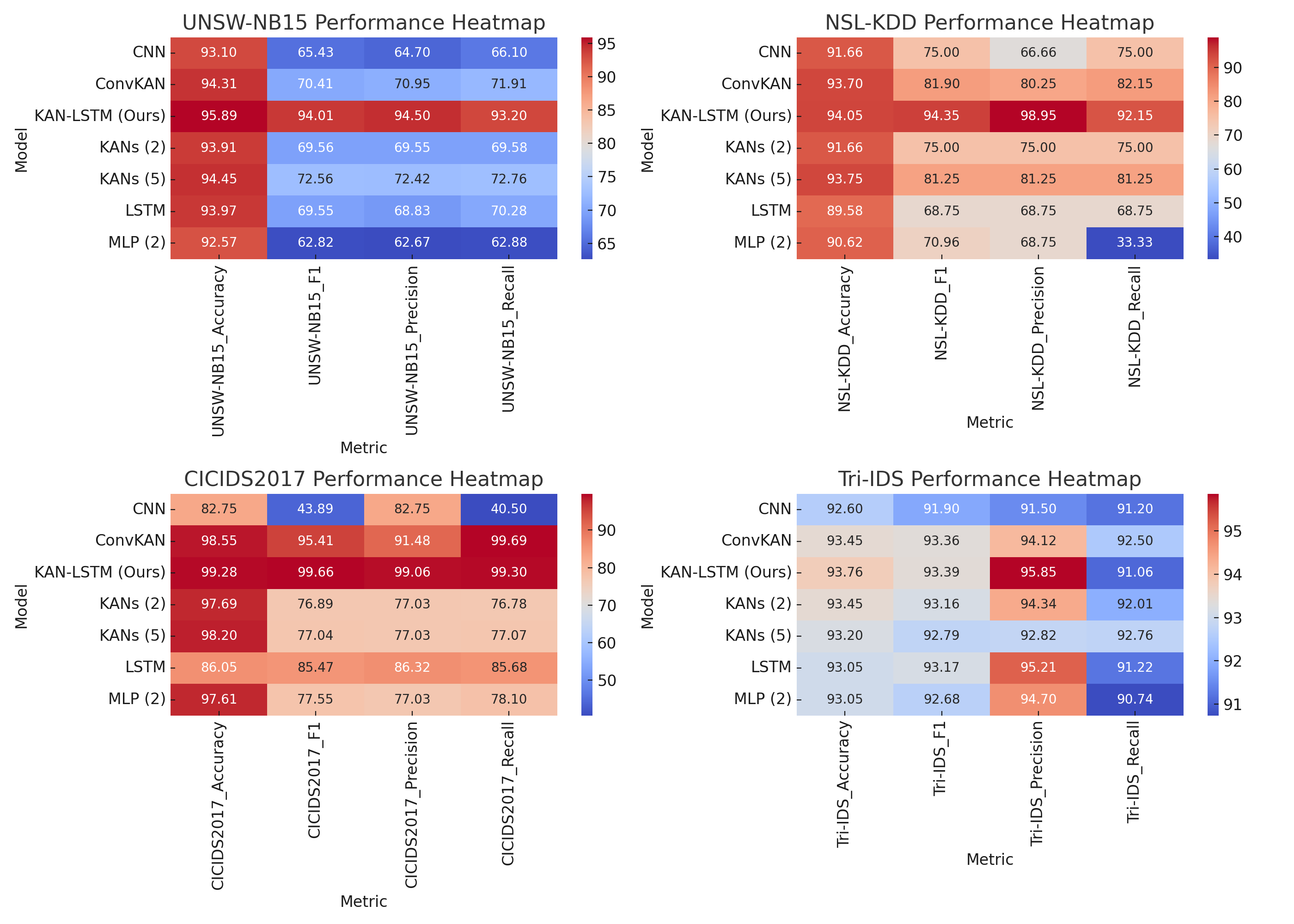}
        \caption{Heatmaps charts.}
        \label{fig:performance_heatmaps}
    \end{subfigure}
    \hfill
    \begin{subfigure}[t]{0.49\textwidth}
        \centering
        % Placeholder for optional fourth figure (or reuse one of the above)
        % Example: You could duplicate grouped bar chart or leave empty
        \includegraphics[width=\linewidth]{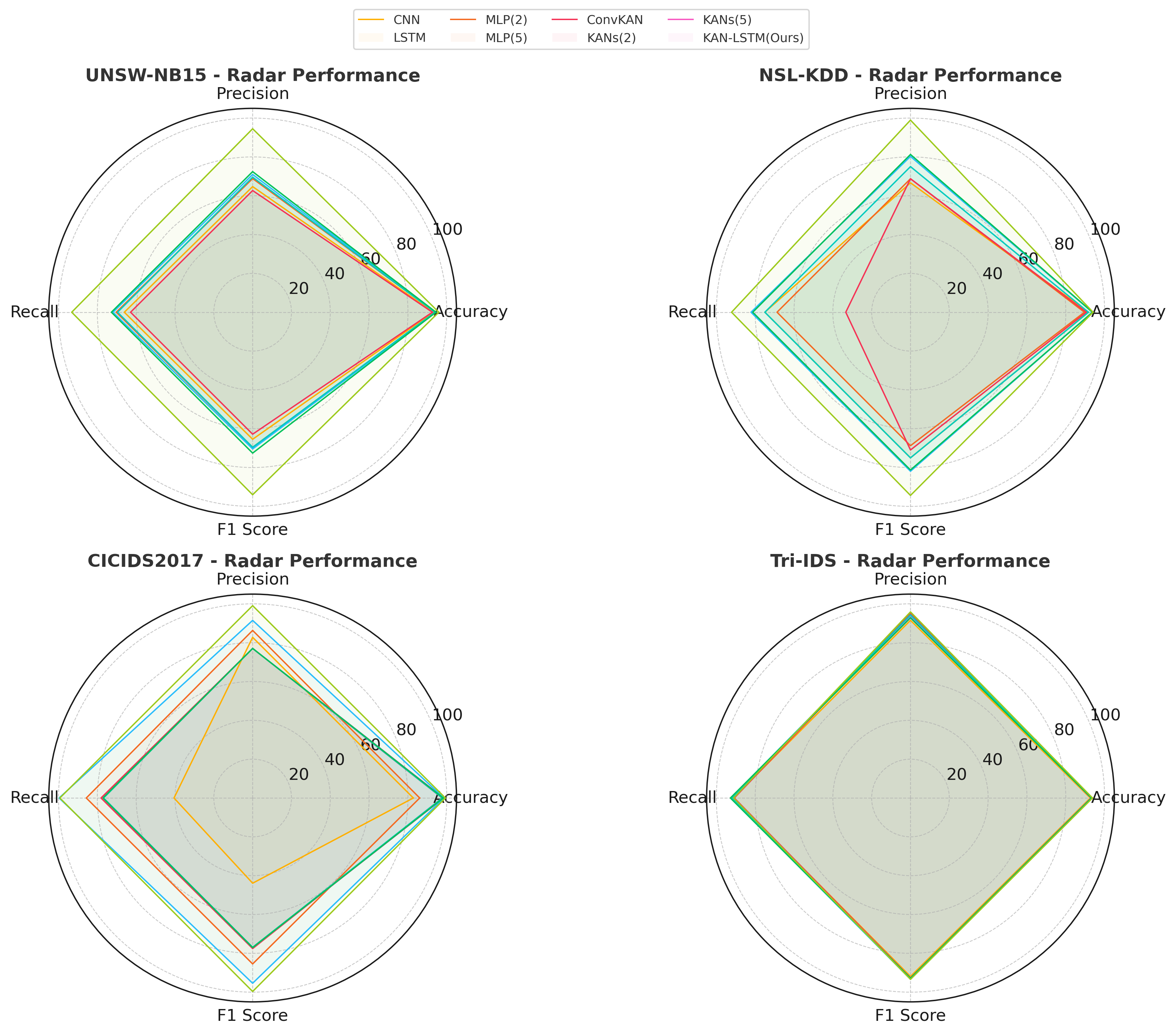}
        \caption{Radar charts.}
        \label{fig:performance_accuracy_only}
    \end{subfigure}

    \caption{
        Visual performance comparison of deep learning models for threat detection across four benchmark datasets
        (UNSW-NB15, NSL-KDD, CICIDS2017, and Tri-IDS) using 
        grouped bar charts, radar charts, and heatmaps. These figures illustrate the superiority of KAN-LSTM over all the other methods. The criteria of  accuracy, precision, recall, and F1 score across datasets are used to show the power of the proposed methods and then KANs, accrodingly.
    }
    \label{fig:model_performance_all}
\end{figure*}

\noindent\textbf{Qualitative Results:}
Four complementary visualizations are used to illustrate the visual comparisons between the baselines on the four datasets,  bar charts, radar plots, heatmaps, and line charts. These  reinforce the quantitative findings from Table~\ref{tab:mytable} and highlight the consistent improvement of the proposed KAN-LSTM obver the rest.

Line Charts (Figure ~\ref{fig:model_performance_all} top left) for each dataset provide the visual comparison in a trend-oriented form. This reveals that while accuracy improvements are relatively modest among top-tier models, the gap in Precision and F1 Score is considerably larger. KAN-LSTM’s curves consistently maintain the upper range across metrics; that is, showing stable performance rather than peaking in a single area. Moreover, the line trajectories of traditional MLP and CNN models are visibly lower and more uneven, which reflects bias and imbalance in the datasets.

Bar Charts (Figure ~\ref{fig:model_performance_all} top right)
visualize all four performance metrics (accuracy, precision, recall, and F1-Score) per dataset in a side-by-side manner for each model. It is clear that while several models exhibit competitive Accuracy (e.g., KANs(5) and ConvKAN), the KAN-LSTM  still achieved higher Precision and F1 Score across datasets. This also illustrates that KAN-LSTM is offering superior balance between detecting malicious traffic and avoiding false positives.

Heatmaps (Figure ~\ref{fig:model_performance_all} bottom left)
offer a quick color-coded comparison, where darker shades indicate stronger performance. Across all datasets, KAN-LSTM cells consistently display the darkest intensities across all four metrics. ConvKAN shows competitive Recall on CICIDS2017, while KANs(5) performs well on certain Accuracy and Precision scores, but neither achieves the top superiority in all the metrics. This illustrates the notion of robustness and adaptability of the proposed model.

Radar Plots (Figure ~\ref{fig:model_performance_all} bottom right)
summarize the multidimensional performance of each model within a single dataset. The proposed KAN-LSTM forms the most outstanding polygon across all datasets which indicates  high scores across all metrics. By contrast, baseline models such as CNN and MLP show irregular, inward shapes that reflect on trade-offs between Precision and Recall. The uniformity of the KAN-LSTM’s radar profile suggests that it does not improve specificity on the cost of  detection sensitivity.

Overall, KAN-LSTM not only outperforms individual baseline models in general but also delivers a balanced metric which is resilient to dataset imbalance and bias. This enables a substantial advantage for intrusion detection tasks in diverse network environments.

\section{Conclusion and future work}
In this study, a comparative evaluations have been conducted on four diverse datasets to illustrate the efficacy and suitability of KANs for cyber security threat and intrusion detection. Three main datasets in the field are chosen, including NSW-NB15, NSL-KDD, and CICIDS2017, as well as a combined dataset of  BOT-IOT + CICIDS2017 + NSL-KDD to provide a diverse, balanced, complex patterns. In order to maintain fairness amongst the comparable baselines, we used the same environment and settings. KANs showed slight performance on the traditional deep learning models such as CNN, MLP, LSTM. Further, we proposed a hybrid architecture to combine LSTM ans KANs which enable the model to capture the diverse patters in the network traffic. Such integration help the model to encode both local relationships and long-range ones. This is the first study to benchmark KANs models to detect intrusions on cyber security threat detection datasets. The empirical results obtained from show the superiority of the proposed model over the baselines with a large margin of 1.5\%. We conclude that this is because of encoding spatial and temporal relationships of the network traffics. However, to achieve the benefits of using KANs in cyber security tasks needs a lot of research to fix the inherent issues such as slow processing, limited scalability and expensive training.  
 \bibliographystyle{elsarticle-num} 
 \bibliography{main.bib}





% \end{thebibliography}
\end{document}